\documentclass[useAMS,usenatbib]{mn2e}

\usepackage{graphicx}
\usepackage{natbib}

\title[A New Ultracool White Dwarf]{A New 
Ultracool White Dwarf Discovered in the SuperCOSMOS Sky Survey}
\author[N. R. Rowell, M. Kilic and N. C. Hambly]{N. R. Rowell$^{1}$\thanks{E-mail:
nr@roe.ac.uk}, M. Kilic$^{2}$ and N. C. Hambly$^{1}$\\
$^{1}$Scottish University's Physics Alliance, Institute for Astronomy, University of Edinburgh, Blackford Hill, Edinburgh  EH9 3HJ\\
$^{2}$The Ohio State University, Department of Astronomy, Columbus, OH 43210}
\begin{document}

\date{Details of acceptance to journal}

\pagerange{\pageref{firstpage}--\pageref{lastpage}} \pubyear{2002}

\maketitle

\label{firstpage}

\begin{abstract}
We present $B_J$, $R_{59F}$ and $I_N$ photometry, and optical and near-infrared spectroscopy, of a new ultracool white 
dwarf (UCWD) discovered in the SuperCOSMOS Sky Survey. The spectrum
of SSSJ1556-0806 shows strong flux suppression due to the presence of collisionally induced absorption
by molecular hydrogen (H2CIA), a feature characteristic of the cool, high density environments found in the atmospheres 
of ultracool white dwarfs. SSSJ1556-0806 therefore joins a list of $<$10 ultracool white dwarfs displaying
extreme flux suppression. 
Synthetic model fitting suggests an effective temperature $<$3000K, which if true would
make this one of the coolest white dwarfs currently known. We also exploit the similarity between the SEDs of SSSJ1556-0806
and the well-studied UCWD LHS 3250 to aid in the determination of the atmospheric parameters in a regime where models consistently
fail to reproduce observations.
SSSJ1556-0806 is relatively bright ($R \sim 17.8$), 
making it particularly amenable to follow up observations to obtain trigonometric parallax and IR photometry.

\end{abstract}


\section{Introduction}
The Galactic disc white dwarf population is of importance in obtaining disc age estimates through
the use of the luminosity function. Now devoid of the nuclear fuel that sustains these stars
during their main sequence lifetime, white dwarfs shine at the expense of their residual thermal
energy, slowly radiating away this fossil heat until they disappear from our telescopes
as cold, compact objects. 
This is a very long process, overlapping the age of the galactic disc at the extreme.
Age determination methods independent of WD cooling rates have been used to estimate the age of the galactic disc at
around 8 Gyr (eg. \citealt{jimenez1998}). \cite{salaris2000} calculate a magnitude of M$_{\textrm{bol}}\sim15.5$ for 0.61M$_{\odot}$
H-rich white dwarfs that have been cooling for 8 Gyr, based on an interpolation of results in their Table 1.
Thus, there is predicted to be a shortfall of white dwarfs at fainter magnitudes, a consequence of the finite age of the
galactic disc population - white dwarfs have simply not had the time to cool further.
The local disk white dwarf luminosity function is thus expected to show a 
marked downturn at luminosities corresponding to cooling ages older than that of the disk.

Ultracool white dwarfs with effective temperatures below $\sim$3700K fall on the cusp of this downturn in 
spatial density. 
They represent some of the oldest stars in the sky, fossils left over from periods of star formation in the very early 
history of the galaxy. Obtaining accurate age estimates
using this method thus depends on finding sufficient numbers of these extremely faint stars to constrain
the observational determination of the luminosity function at the faint end. Large white
dwarf surveys have in the past been hindered by the technical difficulties in finding such a population,
with the limiting magnitude of the survey determining the cut off in the luminosity function, rather than the intrinsic
lack of white dwarfs at cooling ages older than that of the galaxy.

Modern survey class telescopes are now capable of observing the faintest white dwarfs in the disc, and
\cite{liebert1988} were the first to detect the downturn in the luminosity function, at around
M$_{\textrm{bol}}\sim15.76$. \cite{winget1987} used an early version of the observational results to
place a constraint on the age of the galactic disc of $9.3 \pm 2.0$ Gyr. Currently, the largest observational 
determination of the white dwarf luminosity function is that of \citet{harris2006} using SDSS photometry combined
with USNO-B data for improved proper motions. Out of a sample of over six thousand white dwarfs, only four objects 
fell on the downturn in the luminosity function, preventing any quantitative conclusions as to the age of the 
population. Evidently, the reliable identification of faint ultracool white dwarfs, even in small numbers, will enable
tighter constraints to be placed on the spatial density of the oldest white dwarfs, and the statistical uncertainty
in age estimates to be reduced.


\section{Current Observational Status}
In the high pressure atmospheres of cool stars collisions between H$_2$ molecules can induce
temporary dipole moments, allowing these species to absorb radiation \citep{borysow1990}. This so-called H$_2$ 
collisionally induced absorption (H2CIA) introduces broad
absorption features in the infrared, extending into the optical region of the spectrum in the very coolest 
($<$3700K) atmospheres. In mixed atmospheres, collisions between H$_2$ molecules and neutral He can induce the same 
process, and as helium atmospheres are more transparent they are also more dense at the same temperature, with the 
result that collisionally induced absorption becomes important at higher temperatures in mixed atmospheres than for 
pure hydrogen atmospheres. The photospheric opacity of UCWDs is thus dominated by this process,
resulting in spectra showing a strong flux deficit towards the red relative to a blackbody form.

Cool white dwarfs can be roughly divided into two classes. Below around 6000K, opacity due to H2CIA is relatively mild 
and leads to flux suppression in the infrared region of the spectrum. Objects in this category include the white dwarfs
F351-50, LHS 1126 and the well characterised halo white dwarf WD 0346+246 \citep{hambly1997}, for which \cite
{bergeron2001} calculates a T$_{\textrm{eff}}$ of 3780K. With decreasing temperature, H2CIA becomes a more dominant 
source of photospheric opacity.

Below around 3700K, flux suppression extends into the optical region of the SED, shifting the peak \textit
{bluewards} with decreasing T$_{\textrm{eff}}$. This results in a reversal of the slow redwards trend in the optical 
and infrared 
colours of cooling white dwarfs as this process begins to dominate the atmospheric opacity. 
Such ultracool objects are often misidentified by automated
search routines, a fact used by \cite{gates2004} to locate five white dwarfs in the Sloan Digital Sky Survey that show 
varying degrees of flux suppression. Several other UCWDs have been discovered in the SDSS, including SDSS J1337 by 
\citet{harris2001}. These are generally very faint ($r > 19$), making them difficult objects to study in terms
of parallax follow up or IR photometry.

Extreme examples of flux suppression include the stars LHS 1402 and LHS 3250. Observations of the latter are
presented by \citet{oppenheimer2001} alongside those of WD 0346+246, exhibiting the difference between these two
rough groupings.

\section{The SuperCOSMOS Sky Survey}
\label{RPM}
The SuperCOSMOS Sky Survey (SSS) has been compiled by scanning photographic Schmidt plate images taken over a
baseline of up to fifty years, in three different photographic passbands and one twice, giving four 
epochs of observation (see \citealt{SSSI},b,c). The photographic passbands $B_J$, $R_{59F}$ and 
$I_N$ differ from standard Johnson $BRI$, colour transformations are dealt with in section \ref{colours}. The SSS now covers
the entire sky, and with $\sim$5 times more sky coverage than SDSS provides a valuable new resource for identifying
UCWDs.

We have obtained high proper motion candidates ($0.\!\!^{\prime\prime}18$ yr$^{-1} \le \mu_{\textrm{tot}} \le 10.\!\!^{\prime
\prime}00$ yr$^{-1} $) by use of an object pairing algorithm described
in \cite{hambly2004}. White dwarf candidates are selected based on their \textit{Reduced Proper 
Motion} (RPM), an estimate of absolute magnitude that uses proper motion as a distance proxy. A plot of RPM 
against colour is topologically equivalent to the classical HR diagram, and white dwarfs
can be identified in the usual manner by their extreme subluminosity at a given colour. The use of the SSS in defining
a clean, complete sample of white dwarfs based on the technique of RPM is described thoroughly in
\citet{hambly2005}.


\section{A New Ultracool White Dwarf}
We have identified a new ultracool white dwarf in data taken from the SuperCOSMOS Sky Survey. This
star, designated SSSJ1556-0806, shows the strong H2CIA flux deficit associated with the cool, high pressure 
photospheres of white dwarfs of effective temperature $<$3700K.

\subsection{Observations}
\begin{figure}
\resizebox{8cm}{8cm}{
\includegraphics[height=8cm]{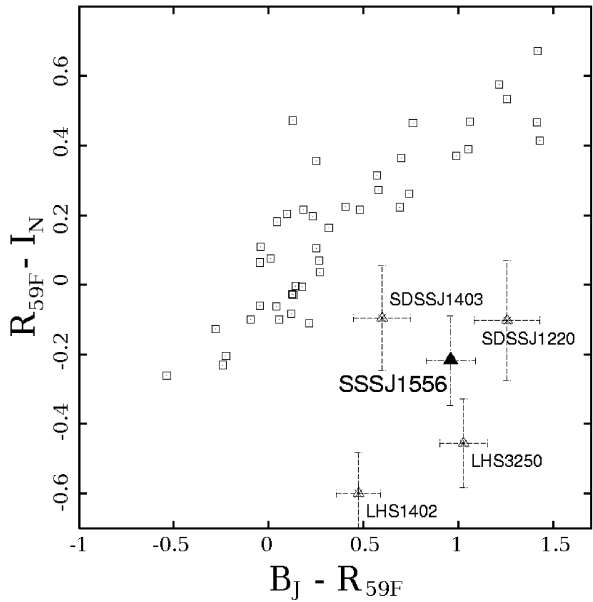}}
\caption{SSSJ1556-0806 lies below the main white dwarf locus in the colour-colour plane, in roughly the same region
as several known ultracool white dwarfs that display extreme flux suppression. Open triangles represent previously 
confirmed UCWDs that appear in SSS. Open squares are normal white dwarfs that appear in SSS
and have been spectroscopically confirmed by \citet{mccook1999}. The filled triangle is SSSJ1556-0806. Error bars have been omitted from
the normal white dwarfs for clarity; they are of order $\sigma_{B-R,R-I}\sim0.1$.}

\label{colour_plot}
\end{figure}

SSSJ1556-0806 was identified as a white dwarf based on it's RPM, as described in
section \ref{RPM}. It was selected as an UCWD candidate
from it's location in the two-colour plane (see Figure \ref{colour_plot}), below the main locus of white dwarfs 
and close to a group of known UCWDs. Various astrometric and photometric data are presented in Table 1.

\begin{table}
\begin{center}
\begin{minipage}{50mm}
  \caption{\scshape Astrometric and Photometric Data for New Ultracool White Dwarf SSSJ1556-0806.}
   \begin{tabular}{@{}lr@{}}

  \hline
Property & Value \\
	\hline
	\hline
Designation & SSSJ1556-0806 \\
Right Ascension\footnotemark[1] & 15 56 47.3181 \\
Declination\footnotemark[1] & -08  5 59.713 \\
Epoch & 1992.480 \\
$\mu_{\textrm{tot}}$ (as yr$^{-1}$) & 0.421$\pm$0.0055 \\
Position Angle & 117.913 \\
$B_J$ & 18.800 \\
$R_{59F}$ & 17.840 \\
$I_N$ & 18.058 \\
 \hline
\end{tabular}
\renewcommand{\thempfootnote}{\arabic{mpfootnote}}
\footnotetext[1]{Coordinates given in equinox 2000}
\end{minipage}
\end{center}
\end{table}


\subsection{Spectroscopy}
\begin{figure}
\begin{minipage}{84mm}
\includegraphics[width = 8.4cm]{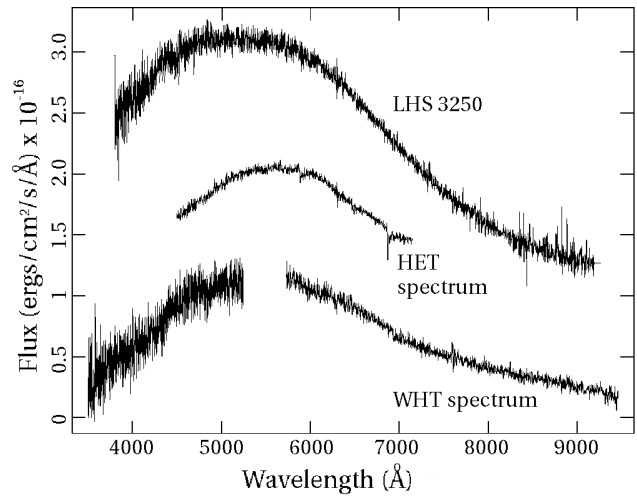}
\caption{Spectra of SSSJ1556-0806 taken using both the Hobby-Eberly telescope and William Herschel telescope. The break 
in the WHT spectra is due to the dichroic beam-splitting mirror used in the dual-arm ISIS spectrograph. The flux scale 
is accurate for the WHT spectrum; the HET spectrum has been offset vertically for clarity. The SDSS spectrum of UCWD
LHS 3250 has been included to demonstrate the similarity between the SEDs of these two objects. }
\label{both_spectra}
\hfill
\end{minipage}
\end{figure}

Follow up spectroscopic observations are necessary to confirm the identity of SSSJ1556+0806 as an UCWD.
These were carried out over two semesters at two different institutions.
Optical spectroscopic observations were made using
the Marcario Low Resolution Spectrograph on the Hobby-Eberly Telescope at McDonald Observatory, Texas, on the 
night of 2006 July 15$^{\textrm{th}}$. Grism 2 in conjunction with a $1.\!\!^{\prime\prime}5$ slit produced 
spectra with a resolution of $\sim$6\AA$ $ over the range 4280-7340\AA.
Optical/near-IR observations were made at the William Herschel Telescope in the Canary Islands on
2007 April 24$^{\textrm{th}}$ using the dual-arm ISIS spectrograph. A $1.\!\!^{\prime\prime}5$ slit was used together with the R300
grating on the blue arm and R158 on the red arm. The resolving power of each arm was, respectively, $\sim$9\AA$ $ and $\sim$11\AA$ $
over the ranges 3160-5270\AA$ $ and 5710-10350\AA.
Data reductions were carried out using standard routines in IRAF.
In Figure \ref{both_spectra} we present the spectra obtained at both of these observing runs, offset vertically for 
clarity. The SDSS spectrum of the well-studied UCWD LHS 3250 has been included (offset vertically) to show the similarity
between the spectral energy distributions of these objects, a feature the will be used to aid analysis in section \ref{LHS3250}.

\section{Atmospheric parameters}

\subsection{The models}
\label{fitting}
In order to derive rough atmospheric parameters for this star, we have fitted the photometric colours to
grids of DA and DB white dwarf synthetic colours computed by P. Bergeron and described in \citet{FBB}. The DA grid 
ranges in effective temperature from 1500K to 100,000K, and the DB grid from 3500K to 30,000K; both cover a range of 
$\log (g)$ from 7.0 to 9.0. A variety of models at fixed masses have also been fitted, these range in mass from 
0.2M$_{\odot}$ to 1.2M$_{\odot}$ in steps of 0.1M$_{\odot}$ and cover a range of T$_{\textrm{eff}}$ and $\log (g)$ 
similar to the previous models. Only pure hydrogen (q$_{\textrm{H}}=10^{-4}$) and pure helium (q$_{\textrm{H}}=10^{-10}
$) atmospheres are considered, mixed atmosphere models have not been used here.

\subsection{Colour transformations}
\label{colours}
The magnitude system used in the SSS is non-standard. The subscripts attached to the magnitudes $B_J$, $R_{59F}$
and $I_N$ refer to particular combinations of colour filter and photographic emulsion, resulting in response
functions different to the standard Johnson $BRI$. In particular, $B_J$ deviates significantly from Johnson $B$, being
closer to a Johnson $V$, and $R_{59F}$ lacks the long red tail that is a characteristic of Johnson $R$.

In order to fit the photographic magnitudes of SSSJ1556-0806 to the photoelectric magnitudes of the models, 
transformation of UKST photographic magnitudes was therefore necessary. For
this we used the colour equations derived by \cite{salim2004} for $B$ and $I$, and \cite{bessel1986} for $R$. Note that other
colour transformations exist, e.g. \cite{blair1982} derive different transformations using all standard
stars in a given region, however Salim et al. use a sample of cool white dwarfs and hence obtain equations more 
suited to our purpose.

\subsection{Fitting procedure}
\label{fitting2}
We have transformed the photographic $B_JR_{59F}I_N$ magnitudes of SSSJ1556-0806 to Johnson magnitudes as 
described previously, and used a variance weighted least squares method to fit these to the photometric colours
of the models. The model predictions are provided in terms of broadband fluxes, hence the fits are done
using these alone rather than the full spectrophotometry.

As the atmosphere content is uncertain, we have found the best fitting DA and DB models for comparison.
The mass distribution of white dwarfs is tightly distributed about the mean, and so for the DA and DB fits
we have restricted the mass to M $=0.603\pm0.081$M$_{\odot}$ and M $=0.718\pm0.111$M$_{\odot}$ respectively, 
values taken from \citet{kepler2007}.

\begin{figure}
\begin{minipage}{84mm}
\includegraphics[width = 8.4cm]{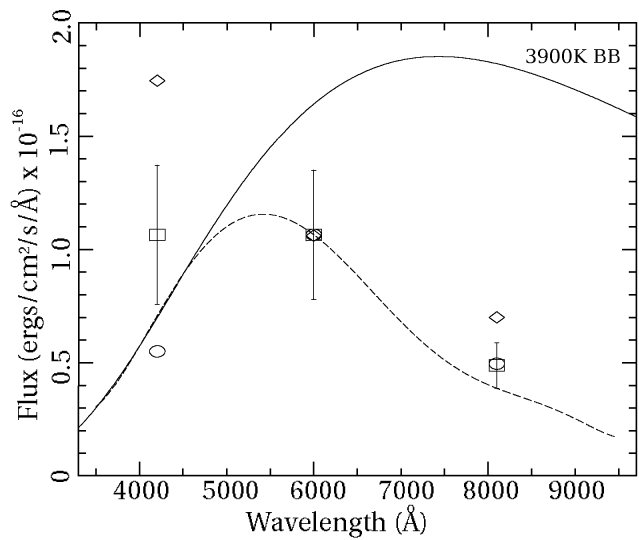}
\caption{Photometric fluxes of the best fitting DA (circles) and DB (diamonds) models are plotted over those of 
SSSJ1556-0806 (boxes). Large error bars are due to a combination of absolute photometric
uncertainty and uncertainty in transformations to Johnson passbands. The dashed line is a ninth order polynomial
fit to the continuum flux of SSSJ1556-0806. Models are provided in terms of broadband fluxes, so our fitting 
procedure is on these alone rather than the full spectrophotometry. 
Model fluxes have been normalised to the $R$ band at 6000\AA. The solid line is a blackbody
fit to the blue end (3500-4800\AA)$ $ of the SSSJ1556-0806 spectrum, the region of the optical spectrum of 
UCWDs that most closely 
approximates a blackbody, and is intended to be roughly illustrative of the effective temperature range.}


\label{fluxes}
\hfill
\end{minipage}
\end{figure}

The parameters of the best fitting models are given in Table 2, and a fit of the model colours to the stellar
colours is presented in Figure \ref{fluxes}, with magnitudes normalised to the R band. These have 
been converted to fluxes, allowing visual comparison with the WHT spectrum, using the prescription outlined below.
Magnitudes in different passbands have been converted into approximate 
average fluxes at central wavelengths using the equation:

\begin{equation}
m = -2.5\log{f^m_{\lambda}} + C_m
\end{equation}
where

\begin{equation}
f^m_{\lambda} = \frac{\int_0^{\infty}f(\lambda) S_m(\lambda) d\lambda}{\int_0^{\infty} S_m(\lambda) d\lambda}
\end{equation}
is the \textit{average} flux in band $m$, $f(\lambda)$ is the flux received from the star at wavelength $\lambda$, and
$S_m(\lambda)$ is the transmission function for bandpass $m$.
The constants $C_m$ have been taken from \citet{BRL1997}, and are calculated from Vega fluxes.

\subsection{Temperature and atmosphere content}

For the DA model, 
the fitted effective temperature of 2500K agrees qualitatively with what one would expect based on visual examination
of the SED - namely, that at this temperature one would expect strong flux suppression extending into the optical region
of the spectrum, and a lack of any discernible atomic absortion lines. The temperature of the DB model is suspiciously
high - even though H2CIA sets in at higher temperatures in mixed atmospheres, at 7000K one would not expect to see such 
strong flux suppression. A cooler temperature is also suggested by fitting a blackbody curve to the blue end - the region of the 
optical spectrum in UCWDs that most closely approximates a blackbody form. Our least-squares fit (in steps of 50K in the range 
3500-4800\AA)$ $ implies an effective temperature $<$4000K, at odds to that predicted by the DB model. Note 
that the
fitting of a blackbody is not in general an accurate way to measure stellar temperature, however in a temperature 
regime where stellar models also fail to accurately reproduce observations (see section \ref{LHS3250}) this can 
at 
least provide an illustrative temperature measurement, and is used frequently in the literature.

The age of the DA model would make this star several Gyr older than the galactic disc, and thus a strong candidate for
a member of the halo population. However, the calculated tangential velocity is not consistent with halo kinematics, 
and would suggest membership of the disc population instead.
The DB model places the star at sufficient distance that the tangential velocity is indicative of halo
membership, and is comparable to that of the classic halo white dwarf WD 0346+246 which has $v_t \sim 170$ kms$^{-1}$.
While the cooling age of $\sim$2.3 Gyr is considerably younger than the halo population, we cannot rule this out as a 
halo star based on this alone. The total stellar age (MS + WD) is required, and this can only be determined with 
accurate distance and mass measurements.

One expects UCWDs to have hydrogen dominated atmospheres due to arguments outlined in \citet{bergeron2001}, 
namely that during their long cooling periods, these stars travel throughout the 
galaxy accreting material from the interstellar medium. While heavier elements tend to quickly sink below the 
photosphere, accreted hydrogen will float on top, resulting in an atmospheric hydrogen abundance that can only increase 
with time. Combined with the observational fact that, at least for hot white dwarfs, DAs outnumber DBs by a factor of 
$\sim$10, we regard the DA model as a more realistic solution. Indeed, the DB model provides a poor fit to the photometric data
and can be ruled out as a likely solution. However, with only two colour indices to fit, it is impossible to draw any firm conclusions as to the 
nature of this object.

\begin{table}
\begin{center}
 \begin{minipage}{80mm}

  \caption{\scshape Atmospheric parameters derived for SSSJ1556-0806 by fitting DA and DB model atmospheres.}
   \begin{tabular}{@{}lrr@{}}
  \hline
 & \multicolumn{2}{c}{Spectral Type}\\
Property & DA & DB \\
	\hline
	\hline
Effective Temperature (K) & $2500$ & $7000$ \\
$\log(g)$ &  $8.04$ & $8.199$ \\
Mass & 0.6 M$_{\odot}$ & 0.7 M$_{\odot}$ \\
Bolometric Correction & $0.364$ & $-0.124$ \\
M$_{\textrm{bol}}$ &  $17.942$ & $13.702$ \\
Age (Gyr) & $12.47$  & $2.315$ \\
M$_{B}$ &  $18.905$ & $14.198$ \\
M$_{R}$ &  $17.023$ & $13.572$ \\
M$_{I}$ &  $17.149$ & $13.320$ \\
Distance (pc)& 12.85 $\pm$ 0.15 & 80.71 $\pm$ 5.52\\
Tangential Velocity (kms$^{-1}$) & $25.64 \pm 0.16$ & $161.06 \pm 8.08$\\
 \hline
\end{tabular}
\end{minipage}
\end{center}
\end{table}
%
%
%
\subsection{Comparison with LHS 3250}
\label{LHS3250}
Synthetic WD spectra fail to accurately reproduce the SEDs of UCWDs, suggesting inadequate or incomplete physics used
in the model atmosphere calculations. In particular, pure hydrogen models show a large absorption feature around
7500\AA$ $ which is not seen in the spectrum of any confirmed UCWD, and the peak of the flux distribution is always predicted 
too narrow. Note that here we have used only model colours, however these are calculated by integrating 
model spectra over the appropriate bandpasses, and are assumed to suffer the same problems.

The similarity between both the SEDs (see Fig.\ref{both_spectra}) and photometric colours ($B_J$ - $R_F$ = 0.96 and 1.028, $R_F$ - $I_N$ = 
-0.218 and -0.456, respectively) of SSSJ1556-0806 and the UCWD LHS 3250 motivates us to consider an alternative approach to the analysis of 
SSSJ1556-0806. LHS 3250 has been studied rigorously - see \citet{harris1999} and \citet{bergeron2002}. Bergeron and Legget fit a variety of 
atmosphere models, covering a large parameter space, to their data, which include trigonometric parallax, $JHK$ photometry and low resolution optical 
spectra. Their
fig.3 demonstrates the disagreement between observed and theoretical SEDs. On this basis they rule out a pure H solution for 
LHS 3250, instead finding a mixed atmosphere solution with $\log$ N(H)/N(He) = -4.7, $\log\textrm{(g)}$ = 8.00 and T$_{\textrm{eff}}$=3480K.

To proceed with our analysis of SSSJ1556-0806, we shall assume that these stars share similar absolute magnitudes and atmospheric properties.
The distance to LHS 3250 is well constrained from trigonometric parallax measurement, allowing us to convert it's SSS apparent magnitudes to absolute 
magnitudes in the photographic $B_JR_{59F}I_N$ passbands.
The apparent magnitudes of SSSJ1556-0806 thus place it at a distance of $31.6\pm1.4$pc, on combining the distances calculated from the individual
passbands. This suggests a tangential velocity of $63.1\pm2.8$kms$^{-1}$, consistent with the galactic thick disc population.

The atmospheric and evolutionary properties are more tricky; Bergeron \& Legget point out that the inconsistency between observed and theoretical 
SEDs rules out
precise determination of these parameters. In the circumstances, we can do no more than say that this star most likely has an effective 
temperature below 4000K and a helium-rich atmosphere. These various properties are summarised in Table 3.
\begin{table}
\begin{center}
\begin{minipage}{60mm}
  \caption{\scshape Atmospheric parameters for SSSJ1556-0806 derived using LHS 3250 as an analogue.}
   \begin{tabular}{@{}lr@{}}

  \hline
Property & Value \\
	\hline
	\hline
Effective Temperature (K) & $<$4000K \\
Atmosphere content & He rich \\
M$_{B_J}$ & 16.302 $\pm$ 0.367 \\
M$_{R_{59F}}$ & 15.274 $\pm$ 0.369 \\
M$_{I_N}$ & 15.730 $\pm$ 0.392 \\
Distance (pc) & 31.6 $\pm$ 1.4 \\
Tangential Velocity (kms$^{-1}$) & 63.1 $\pm$ 2.8 \\
 \hline
\end{tabular}
\end{minipage}
\end{center}
\end{table}
\section{Conclusion}
We have presented observations of a new ultracool white dwarf found in the SuperCOSMOS Sky Survey, and made
an initial effort to understand the unusual SED of this object using both atmosphere models for cool white dwarfs
and comparison to the well-studied UCWD LHS 3250.
Without better photometry, models are poorly constrained and can do no more than give a rough interpretation of the object.
The current inconsistencies between the theoretical and observed spectral energy distributions of UCWDs rules out any precise determination
of the atmospheric parameters. The most likely interpretation of this object is as a thick disk white dwarf, with a helium dominated
atmosphere and an effective temperature $<$4000K.


Infrared photometry is essential for a good fit to theoretical models. Trigonometric parallax observations can also
provide constraints by measuring both the absolute magnitude and, indirectly, mass. 
We currently have time allocated on the robotic Liverpool Telescope for a program to obtain parallax
measurements of UCWDs. SSSJ1556-0806 has been added to our queue-scheduled list and has already had one epoch of 
observation carried out at high parallax factor.


\bibliographystyle{mn2e}
\bibliography{references}

\bsp

\label{lastpage}

\end{document}